\documentclass[preprint,sort&compress]{elsarticle}
\usepackage{amsmath,amssymb,latexsym}
\usepackage{bm}
\usepackage{braket}
\usepackage{graphicx}
\usepackage{hepunits}
\usepackage[svgnames]{xcolor}
\usepackage{hyperref}

\journal{Physics Letters B}
\newcommand{\w}{\omega}

\allowdisplaybreaks

\begin{document}

\begin{frontmatter}

\title{Constructing Canonical Feynman Integrals with Intersection Theory}

\author[1]{Jiaqi Chen}
\ead{jiaqichen@pku.edu.cn}
\author[1]{Xuhang Jiang}
\author[2]{Xiaofeng Xu}
\ead{pkuxxf@gmail.com}
\author[3]{Li Lin Yang}
\ead{yanglilin@zju.edu.cn}
\address[1]{School of Physics and State Key Laboratory of Nuclear Physics and Technology, Peking University, Beijing 100871, China}
\address[2]{Institut f\"ur Theoretische Physik, Universit\"at Bern, Sidlerstrasse 5, CH-3012 Bern, Switzerland}
\address[3]{Zhejiang Institute of Modern Physics, Department of Physics, Zhejiang University, Hangzhou 310027, China}

\begin{abstract}

Canonical Feynman integrals are of great interest in the study of scattering amplitudes at the multi-loop level. We propose to construct $d\log$-form integrals of the hypergeometric type, treat them as a representation of Feynman integrals, and project them into master integrals using intersection theory. This provides a constructive way to build canonical master integrals whose differential equations can be solved easily. We use our method to investigate both the maximally cut integrals and the uncut ones at one and two loops, and demonstrate its applicability in problems with multiple scales.

\end{abstract}

\end{frontmatter}


\section{Introduction}

Functions of uniform transcendentality (UT) are of great interest in the studies of scattering amplitudes in quantum field theories. They admit series expansions of the form
$
f(\epsilon, \vec{x}) = \sum_{n=0}^\infty \epsilon^n f^{(n)}(\vec{x})
$,
where $\epsilon$ is the dimensional regulator and $f^{(n)}(\vec{x})$ is a function of transcendental weight $n$.
When calculating loop amplitudes for scattering processes, it is extremely useful to find a basis of Feynman integrals consisting of UT functions whenever possible. Such Feynman integrals are dubbed as the ``canonical'' ones. Canonical Feynman integrals satisfy differential equations of the $\epsilon$-form \cite{Henn:2013pwa}
$
df_i(\epsilon, \vec{x}) = \epsilon \, dA_{ij}(\vec{x}) f_j(\epsilon, \vec{x})
$,
where $A$ is an algebraic matrix of the variables $\vec{x}$. This kind of equations can be solved order-by-order in $\epsilon$ as iterated integrals \cite{Chen:1977oja}, often leading to compact analytic expressions in terms of (multiple) polylogarithms \cite{Goncharov:1998kja} that allow efficient numerical evaluation~\cite{Vollinga:2004sn}. Even in cases when explicit analytic expressions (at high weights) are difficult to find, the iterated integrals can be easily performed via numeric integration \cite{Caron-Huot:2014lda, Bonciani:2016qxi, Becchetti:2017abb, Xu:2018eos, Wang:2019fxh} or series expansion \cite{Lee:2017qql, Bonciani:2019jyb, Frellesvig:2019byn, Francesco:2019yqt, Hidding:2020ytt}.

Due to the importance of canonical bases, various methods have been proposed in the literature to transform a given set of differential equations into the $\epsilon$-form \cite{Henn:2013pwa, Argeri:2014qva, Lee:2014ioa, Gehrmann:2014bfa, Georgoudis:2016wff, Becchetti:2017abb, Dlapa:2020cwj}. Certain algorithms have been implemented into public program packages \cite{Georgoudis:2016wff, Gituliar:2017vzm, Prausa:2017ltv, Meyer:2017joq}. They have been successfully applied to many multi-loop calculations. However, when the number of mass scales increases, direct application of such automated algorithms often becomes inadequate due to the appearance of many irrational functions (square roots) which cannot be simultaneously rationalized. In these cases (e.g., \cite{Gehrmann:2014bfa, Becchetti:2017abb, Xu:2018eos, Wang:2019fxh, Heller:2019gkq, Chen:2019zoy, Becchetti:2019tjy}), manual intervention is usually required to achieve the goal.

It has been realized that canonical Feynman integrals are closely related to $d\log$-form integrals \cite{ArkaniHamed:2010gh, Gehrmann:2011xn, Drummond:2013nda, Arkani-Hamed:2013jha, Arkani-Hamed:2014via, Bern:2014kca, Arkani-Hamed:2016byb, Chicherin:2018old, Herrmann:2019upk, Henn:2020lye}. 
They lead to beautiful geometric pictures for the scattering amplitudes in planar $\mathcal{N}=4$ supersymmetric theories.
In \cite{Henn:2020lye}, an algorithm to find $d\log$-form integrands in the momentum space has been proposed. Given a set of denominators, their methods make a generic ansatz for the numerator with a couple of to-be-determined coefficient functions, and search for possible forms of the coefficient functions giving $d\log$-form integrals.

In this work, we approach the problem from a different perspective. Instead of manipulating loop integrals directly, we look for generic hypergeometric integrals~\cite{Aomoto:1414035} which \emph{can} have an interpretation as Feynman integrals. Our method does not require making ans\"atze, and is a direct construction starting from a multi-valued function determined by the specific integral topology under consideration. We build all possible $d\log$-form integrals which can be interpreted as Feynman integrals in this topology, and project them back to loop integrals using the intersection theory~\cite{Mizera:2017rqa, Mastrolia:2018uzb, Frellesvig:2019kgj, Frellesvig:2019uqt, Mizera:2019vvs, Weinzierl:2020xyy, Frellesvig:2020qot}. This amounts to exploiting the geometric picture of the hypergeometric integrals, and computing the ``inner-products'' of them using concepts from algebraic geometry. We will use the Baikov representation of Feynman integrals~\cite{Baikov:1996iu} as the concrete prototype to demonstrate our method, but our approach is not confined to that. Our method serves as a constructive way to find canonical Feynman integrals without analyzing the differential equations.

\section{Canonical Feynman Integrals in the Baikov representation}
\label{sec:baikov}

We consider $L$-loop Feynman integrals with $E+1$ external legs in spacetime dimension $d=4-2\epsilon$. The loop momenta are labelled by $k_i$ ($i=1,\ldots,L$) and the independent external momenta are $p_i$ ($i=1,\ldots,E$). For later convenience we collectively refer to them as $q_i$ ($i=1,\ldots,L+M$), where $q_i \equiv k_i$ ($i=1,\ldots,L$), and $q_{L+i} \equiv p_i$ ($i=1,\ldots,E$).
Out of these momenta one can construct $N \equiv L(L+1)/2 + LE$ independent scalar products involving at least one of the $k_i$. An integral family is then defined by a given set of $N$ independent propagators $D_i$ ($i=1,\ldots,N$), which are linear functions of the aforementioned scalar products. A generic integral in such a family is given by
\begin{equation}
F_{a_1,\ldots,a_N} = e^{\epsilon \gamma_E L} \int \bigg[ \prod_{i=1}^{L}  \frac{d^d k_i}{i \pi^{d/2}} \bigg] \frac{1}{D_1^{a_1} \, D_2^{a_2} \cdots D_N^{a_N}} \, ,
\label{eq:loop-int}
\end{equation}
where $a_i \in \mathbb{Z}$ and $\gamma_E$ is the Euler constant. A specific topology in the integral family is defined by a chosen subset of the powers $\{a_i\}$ whose values are positive, while the other powers are either zero or negative.

The Baikov representation of the above integral amounts to a change of integration variables from the set $\{k_i^\mu\}$ to the set $\{D_i\}$. The $\{D_i\}$-independent degrees of freedom can be integrated out giving rise to the Gram determinant $G(\{D_i\}) \equiv \det(q_i \cdot q_j)$. We will use the loop-by-loop construction where one performs the change of variables for a single loop momentum at a time, treating the others as external. The resulting Baikov representation can be written as
\begin{equation}
F_{a_1,\ldots,a_N} = \mathcal{N}_\epsilon \int_{\mathcal{C}} \bigg[ \prod_{i} \big[ G_i(\bm{z}) \big]^{-\gamma_i-\beta_i\epsilon} \bigg] \prod_{j=1}^n \frac{dz_j}{z_j^{\alpha_j}} \, ,
\label{eq:baikov_lbl}
\end{equation}
where $\bm{z} \equiv \{z_1,\ldots,z_n\}$ is a subset of $\{D_i\}$ containing those propagators appearing in the construction, $\{\alpha_j\}$ is the corresponding subset of $\{a_i\}$, $\beta_i \in \mathbb{Z}$ and $\gamma_i$ can be integer or half-integer.
The prefactor $\mathcal{N}_\epsilon$ is a function of $\epsilon$ only. It is not relevant for our discussions and we will often drop it in the following. The integration domain $\mathcal{C}$ is given by the interior of the contour where $G_i(\bm{z})$ vanishes.

We will be concerned with (linear combinations of) Feynman integrals which are canonical. That is, when multiplied by suitable (and easy to find) overall factors, they become UT functions. Our starting point is (generalized) $d\log$-form integrals of the form
\begin{equation}
\int_{\mathcal{C}} \bigg[ \prod_{i} \big[ {G}_i(\bm{z}) \big]^{-\beta_i\epsilon} \bigg] \prod_{j=1}^n d\log f_j(\bm{z}) \, ,
\label{eq:baikov_dlog}
\end{equation}
where $f_j(\bm{z})$ are algebraic functions of the Baikov variables. These integrals apparently lead to UT functions, but it is not \textit{a priori} clear how they are related to Feynman integrals and whether all master integrals in a canonical basis (when it exists) can be represented in this form. In this work, we are precisely dealing with this problem: given an integral family, we'd like to construct as many as possible linearly-independent $d\log$-form integrals, and convert them to linear combinations of Feynman integrals. These then serve as candidates for canonical master integrals from which one can derive differential equations of the $\epsilon$-form. 
The construction of $d\log$-form integrals is the main theme of this work.
Before going into that, in the next section, we first address the problem of converting them to Feynman integrals using the intersection theory~\cite{Mizera:2017rqa, Mastrolia:2018uzb, Frellesvig:2019kgj, Frellesvig:2019uqt, Mizera:2019vvs, Weinzierl:2020xyy, Frellesvig:2020qot}.

\section{The intersection theory for hypergeometric integrals}
\label{sec:intersection}

The integrals in the form of Eq.~\eqref{eq:baikov_lbl} or Eq.~\eqref{eq:baikov_dlog} have a natural interpretation in the geometric language of hypergeometric functions \cite{Aomoto:1414035}. They are defined as
\begin{equation}
\int_{\mathcal{C}} u(\bm{z}) \, \varphi(\bm{z}) \, ,
\label{eq:hyperint}
\end{equation}
where $\bm{z}=(z_1,\ldots,z_n)$, $u(\bm{z})$ is a multi-valued function of $\bm{z}$, and $\varphi(\bm{z})$ is a single-valued differential $n$-form $\varphi(\bm{z}) \equiv \hat{\varphi}(\bm{z}) d^n\bm{z}$. It is assumed that the function $u(\bm{z})$ vanishes on the boundary of the integration domain $\mathcal{C}$. For integrals in Eqs.~\eqref{eq:baikov_lbl} and \eqref{eq:baikov_dlog}, it is straightforward to identify
\begin{equation}
u(\bm{z}) = \prod_{i} \big[ {G}_i(\bm{z}) \big]^{-\gamma_i-\beta_i\epsilon} \, .
\label{eq:uz}
\end{equation}
The $n$-form $\varphi(\bm{z})$ corresponding to Eq.~\eqref{eq:baikov_dlog} is then
\begin{equation}
\varphi(\bm{z}) = \bigg[\prod_{i} \big[ G_i(\bm{z}) \big]^{\gamma_i} \bigg] \bigwedge_{j} d\log f_j(\bm{z}) \, .
\label{eq:phiz}
\end{equation}
It should be stressed that $\gamma_i$ can be half-integers. In this case the functions $f_j(\bm{z})$ must be carefully chosen to ensure that $\varphi(\bm{z})$ is single-valued, such that the integral belongs to an equivalence class of the Feynman integrals in Eq.~\eqref{eq:baikov_lbl}.

The integral Eq.~\eqref{eq:hyperint} is invariant under a gauge transformation of $\varphi(\bm{z})$:
$
\varphi(\bm{z}) \to \varphi(\bm{z}) + \nabla_\w \xi
$,
where $\xi$ is an $(n-1)$-form, $\nabla_\w \equiv d + \w \wedge$ is the covariant derivative with the connection $w(\bm{z}) \equiv d\log(u(\bm{z}))$. This then defines an equivalence class
$
\bra{\varphi} : \varphi \sim \varphi + \nabla_\w \xi
$,
which can be regarded as a twisted cocycle in the twisted cohomology group $H_\w^n$. 

The strategy to convert $d\log$-form integrals (as Baikov representations) to Feynman integrals then proceeds as follows. We choose an arbitrary basis of Feynman integrals for the integral family, derive their Baikov representations using the loop-by-loop construction, and denote their corresponding cocycles as $\{\bra{e_i}\}$. These cocycles then form a basis of the vector space $H_\w^n$, such that the cocycle defined by Eq.~\eqref{eq:phiz} can be written as their linear combination, i.e.,
$
\bra{\varphi} = \sum_{i} c_i \bra{e_i}
$.
Therefore, once we know the coefficients $c_i$, we will be able to write the Baikov representation Eq.~\eqref{eq:baikov_dlog} as a linear combination of Feynman integrals.

The decomposition coefficient $c_i$ can be calculated by considering the dual twisted cohomology $(H_\w^n)^* = H_{-\w}^n$. Choosing a basis of the dual space as $\ket{h_i}$, the coefficients are given by
\begin{equation}
c_i= \sum_j \braket{\varphi | h_j} (\bm{C}^{-1})_{ji}  \, , \quad \bm{C}_{ij} = \braket{e_i | h_j} \, ,
\end{equation}
where $\braket{\varphi_L | \varphi_R}$ is the inner-product of the cocycle $\bra{\varphi_L}$ and the dual vector $\ket{\varphi_R}$ and is called an intersection number. Algorithms for computing the intersection numbers were proposed in \cite{Frellesvig:2019kgj, Frellesvig:2019uqt, Weinzierl:2020xyy, Frellesvig:2020qot}.

\section{Constructing canonical integrals: the univariate case}
\label{sec:univar}

We first study the case where the integrand of Eq.~\eqref{eq:hyperint} lives on a (complex) 1-dimensional manifold, and hence there is only one variable $z$. This serves as a primary step towards the generic multivariate case. The 1-dimensional manifold can also be regarded as the sub-manifold of a higher dimensional manifold. This is applicable, e.g., when considering the maximally cut integrals in the Baikov representation \cite{Primo:2016ebd, Primo:2017ipr, Frellesvig:2017aai}.

The univariate integrals take the form of Eq.~\eqref{eq:hyperint} where the collection $\bm{z}$ contains only a single variable $z$, with $u(z)$ given by Eq.~\eqref{eq:uz}.
We now need to construct possible single-valued 1-forms $\phi(z)$ in the form of Eq.~\eqref{eq:phiz}. Note that depending on the values of $\gamma_i$'s, it is not always possible to find such 1-forms giving rise to $d\log$ integrals. In particular, if more than one $\gamma_i$'s are half-integers, or if some $\gamma_i$ is a half-integer and the corresponding polynomial $G_i(z)$ has more than two distinct roots, the integral is an elliptic integral and is beyond the scope of the current work. Therefore we only need to consider two cases: 1) all $\gamma_i$'s are integers; and 2) there is exact one half-integer $\gamma_i$ and the corresponding $G_i(z)$ has two (or fewer) distinct roots.

In the case when all $\gamma_i$'s are integers, one can always factorize $u(z)$ into the form
\begin{equation}
u(z) = \frac{\mathcal{K}_1^{\epsilon}}{\mathcal{K}_0} \prod_{j=0}^\nu (z-c_j)^{-\gamma'_j-\beta'_j\epsilon} \, ,
\end{equation}
where $\mathcal{K}_0$ is an algebraic function and $\mathcal{K}_1$ is a rational function of the external momenta, respectively;
$c_j$ is a root of one of the polynomials $G_i(z)$ in Eq.~\eqref{eq:uz}; $\gamma'_j$ and $\beta'_j$ are integers. The connection $\w=d\log(u)$ has $\nu$ critical points where $\w = 0$, which means that there exist $\nu$ independent integrals \cite{Lee:2013hzt, Frellesvig:2019uqt}. We can construct $\nu$ nonequivalent 1-forms $\phi(z)=\hat{\phi}(z)dz$ with
\begin{align}
\hat{\phi}_i(z) = \frac{\mathcal{K}_0}{z-c_i} \prod_{j=0}^{\nu}(z-c_j)^{\gamma'_j} \, , \; (i=1,\ldots,\nu) \, ,
\label{eq:phiz_integer}
\end{align}
which give the canonical basis we desired.

On the other hand, if one of the $\gamma_i$'s is a half-integer, without loss of generality, we may write
\begin{equation}
u(z) =\frac{\mathcal{K}_1^{\epsilon}}{\mathcal{K}_0} \big[ (z-c_0)(z-c_1) \big]^{-\gamma_1-\beta_1\epsilon} \prod_{j=2}^\nu (z-c_j)^{-\gamma'_j-\beta'_j\epsilon} \, ,
\label{eq:u_half_integer}
\end{equation}
where $\gamma_1$ is a half-integer.
Again the connection $w=d\log(u)$ has $\nu$ critical points and we need to construct $\nu$ $d\log$-form integrals. For that we use the identities
\begin{align}
\frac{\partial}{\partial x} \log \frac{1+\sqrt{\frac{(x_2-c)(x_1-x)}{(x_1-c)(x_2-x)}}}{1-\sqrt{\frac{(x_2-c)(x_1-x)}{(x_1-c) (x_2-x)}}}
&=
\frac{\sqrt{(x_1-c)(x_2-c)}}{(x-c)\sqrt{(x-x_1)(x-x_2)}} \, , \nonumber
\\
\frac{\partial}{\partial x} \log \frac{1+\sqrt{\frac{(x_1-x)}{(x_2-x)}}}{1-\sqrt{\frac{(x_1-x)}{(x_2-x)}}} &= \frac{1}{\sqrt{(x-x_1)(x-x_2)}} \, ,
\label{eq:deq_fi_sqrt}
\end{align}
up to irrelevant phases. We can then construct the following $\hat{\phi}(z)$:
\begin{align}
\hat{\phi}_1(z) &= \frac{\mathcal{K}_0}{\big[ (z-c_0)(z-c_1) \big]^{1/2-\gamma_1}} \prod_{j=2}^\nu (z-c_j)^{\gamma'_j} \, ,
\label{eq:phiz_half_integer}
\\
\hat{\phi}_i(z) &= \frac{\mathcal{K}_0}{z-c_i} \frac{\sqrt{(c_0-c_i)(c_1-c_i)}}{\big[ (z-c_0)(z-c_1) \big]^{1/2-\gamma_1}} \prod_{j=2}^\nu (z-c_j)^{\gamma'_j} \, , \nonumber
\end{align}
where $i=2,\ldots,\nu$.

It is instructive to see how the above generic $d\log$-form integrals look like in practice, and how they can be related to Feynman integrals. For that we use the two-loop four-scale triangle integrals from \cite{Wang:2019fxh} as a concrete example in the following. More examples can be found in the Supplemental Materials.

\begin{figure}[t!]
\centering
\includegraphics[width=0.4\textwidth]{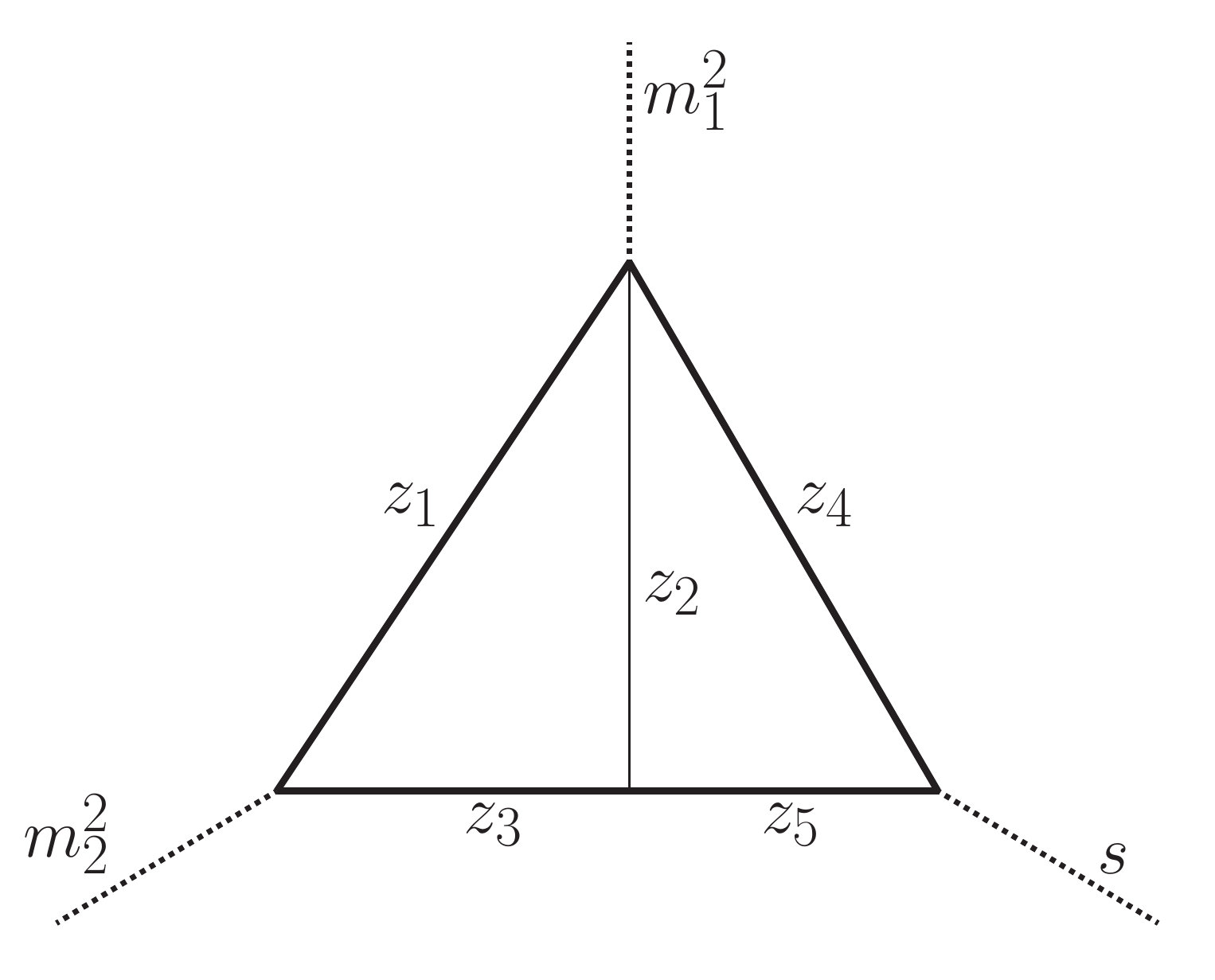}
\caption{\label{fig:hzz}The integral topology considered as an example in the text.}
\end{figure}

The integral family is defined by the propagators
\begin{align}
&\{k_1^2-m^2, \, (k_1-k_2)^2, \, (k_1+p_2)^2-m^2, \, (k_2-p_1)^2-m^2, \nonumber
\\
&(k_2+p_2)^2-m^2, \, z \equiv z_6 \equiv k_2^2-m^2, \, (k_1-p_1)^2-m^2 \} \, ,
\end{align}
where the external momenta satisfy $p_1^2=m_1^2$, $p_2^2=m_2^2$ and $(p_1+p_2)^2=s$. We consider integrals in the sector $\{1,1,1,1,1,0,0\}$, as depicted in Fig.~\ref{fig:hzz}. The 6th propagator $z \equiv z_6$ is an irreducible scalar product (ISP) for constructing the Baikov representation for this topology. After imposing the maximal cuts, the corresponding $u(z)$ is given by
\begin{equation}
u(z) = \frac{1}{\sqrt{-\lambda}} \, z^{-2\epsilon} \left[ -(z-c_0) (z-c_1) \right]^{-1/2+\epsilon}
\big[ (z-c_2) (z-c_3) \big]^{-\epsilon}  \, ,
\end{equation}
where the 4 roots are
\begin{align}
c_{0,1} &= m_2 (m_2 \pm 2m) \, , \nonumber
\\
c_{2,3} &= \frac{s (m_1^2 + m_2^2 - s) \pm \sqrt{s (s-4m^2) \lambda}}{2s} \, ,
\end{align}
with $\lambda \equiv \lambda(s,m_1^2,m_2^2)$ being the K\"all\'en function
\begin{equation}
\lambda(x,y,z) = x^2 + y^2 + z^2 - 2xy - 2yz - 2zx \, .
\end{equation}

The connection $\w = d\log(u)$ has 4 critical points and we need to construct 4 independent canonical master integrals. According to Eq.~\eqref{eq:phiz_half_integer}, we have
\begin{align}
\hat{\phi}_1(z) &= \sqrt{\lambda}  \, , \quad \hat{\phi}_4(z) = \sqrt{\lambda}  \frac{\sqrt{c_0c_1}}{z}  \, , \nonumber
\\
\hat{\phi}_{2,3}(z) &= \sqrt{\lambda} \frac{\sqrt{(c_0-c_{2,3})(c_1-c_{2,3})}}{z-c_{2,3}}  \, .
\end{align}
We now need to convert these 1-forms to (maximally cut) Feynman integrals. For that we choose the basis 
$F_{1,1,1,1,1,0,0}$, $ F_{2,1,1,1,1,0,0}$, $F_{2,1,1,2,1,0,0}$ and $ F_{1,1,1,2,1,0,0}$. Their corresponding cocycles are $\bra{e_i}$ with $i=1,\ldots,4$.
Computing the intersection numbers, we then have
\begin{align}
\bra{\phi_1} &=\sqrt{\lambda} \bra{e_1} \, , \quad \bra{\phi_4} = \frac{\sqrt{\lambda}\sqrt{m_2^2(m_2^2-4 m^2)}}{2 \epsilon} \bra{e_2} \, , \nonumber
\\
\bra{\phi_{2,3}} &= \frac{s(s-m_1^2-m_2^2)}{2\epsilon} \bra{e_4} - \frac{m_2^2(s+m_1^2-m_2^2)}{2\epsilon} \bra{e_2} + \frac{\lambda m^2 + s m_1^2 m_2^2}{2\epsilon^2} \bra{e_3} \nonumber
\\
&\mp \frac{\sqrt{\lambda} \sqrt{s(s-4m^2)}}{2 \epsilon}  \bra{e_4} \, .
\end{align}
We have checked that the homogeneous part of their differential equations indeed takes the $\epsilon$-form.

\section{Constructing canonical integrals: the multivariate case}
\label{sec:mulvar}

We now turn to the generic multivariate case. Our strategy is to construct for one variable at a time, using the building blocks presented in the previous section. Our starting point is again Eqs.~\eqref{eq:hyperint} and \eqref{eq:uz}. We pick a variable which satisfies the criteria outlined in the second paragraph of the last section. Without loss of generality, we will call this variable $z_1$. Using the method from the previous section, we can find a collection of functions $\hat{\varphi}^{(1)}_i(\bm{z})$ such that
\begin{equation}
\hat{\varphi}^{(1)}_i(\bm{z}) = \bigg[\prod_{j} \big[ G_j(\bm{z}) \big]^{\gamma_j} \bigg] \frac{\partial}{\partial z_1} \log f^{(1)}_i(\bm{z}) \, .
\label{eq:phi1z}
\end{equation}
We denote the combination $u(\bm{z}) \hat{\varphi}^{(1)}_i(\bm{z}) dz_1$ as a partial-$d\log$-form integrand. Note that $d\log f_1(\bm{z})$ actually produces all $dz_i$ terms for $i > 1$. However, they always give vanishing results when taking the wedge product with the other factors to be determined later. We also note that under the square roots in $f^{(1)}_i(\bm{z})$, there can be cubic or quartic powers of $z_i$ for $i > 1$, which nevertheless does not pose a problem at this step.

We now need to pick the next variable $z_2$ and repeat the above procedure. This amounts to construct the second-level coefficients $\hat{\varphi}^{(2)}_{i,j}(\bm{z}')$ in the sense of multivariate intersection theory \cite{Frellesvig:2019uqt, Weinzierl:2020xyy, Frellesvig:2020qot}, where $\bm{z}'=\{z_2,\ldots,z_n\}$. The goal is to make $u(\bm{z}) \hat{\varphi}^{(1)}_j(\bm{z}) dz_1 \wedge \hat{\varphi}^{(2)}_{i,j}(\bm{z}') dz_2$ a partial-$d\log$-form integrand in two variables. Such a recursive procedure leads to the final results for the $d\log$-form integrals in this topology.

There is, however, one subtlety at this point. While $\hat{\varphi}^{(1)}_j(\bm{z})$ are rational functions of $z_1$ by construction, they often involve square roots of $z_2$-polynomials. It is then necessary that $\hat{\varphi}^{(2)}_{i,j}(\bm{z}')$ are also algebraic functions of $z_2$, which make the product $\hat{\varphi}^{(1)}_j(\bm{z}) \hat{\varphi}^{(2)}_{i,j}(\bm{z}')$ rational functions of both $z_1$ and $z_2$. In the same time, one needs to invoke Eq.~\eqref{eq:deq_fi_sqrt} in the construction which requires that $\hat{\varphi}^{(2)}_{i,j}(\bm{z}')$ should have square roots of $z_2$-polynomials as overall factors. Therefore we use the properties of Gram determinants to make linear combinations of the solutions at the $z_1$-level, such that $\hat{\varphi}^{(1)}_j(\bm{z})$ takes the form $\sqrt{\Lambda(\bm{z}')} \, g(\bm{z})$, where both $\Lambda(\bm{z}')$ and $g(\bm{z})$ are rational functions. The construction can then proceed given that $\Lambda(\bm{z}')$ is a quadratic polynomial of $z_2$. Note that even if the square roots in $u(\bm{z})$ involves higher powers of $z_2$, they are not necessarily present in $\Lambda(\bm{z}')$.

During the recursive procedure outlined above, it may happen that none of the remaining variables in $\bm{z}'$ satisfies the criteria of our construction. 
In this case, one may try to start from a different parametrization (e.g., changing variables, or using the standard Baikov representations instead of the loop-by-loop ones) which gives a different $u(\bm{z})$ function.
If no such parametrization could be found, it is likely that the canonical basis does not exist from the beginning, although we cannot exclude the case where a UT function does not admit a $d\log$-form integral representation in terms of Baikov variables.

Given the above general idea, it is best to see it in action. We again use the two-loop four-scale triangle integrals as an example. We pick $z=z_6$ as the first variable to construct. The complete $u(\bm{z})$ function without cuts is given by
\begin{equation}
u(\bm{z})=\frac{(-\lambda)^{-1/2} \, s^{-\epsilon} (z_3+m^2)^{-\epsilon}}{\left[ -(z-c_0)(z-c_1) \right]^{1/2-\epsilon}} \prod_{i=2}^{5} (z-c_i)^{-\epsilon} \, ,
\end{equation}
where $\bm{z} = \{z,z_1,\ldots,z_5\}$. To demonstrate the idea, it is enough to have
\begin{align}
c_{0,1} &= m_2^2 + z_5 \pm 2 \sqrt{m_2^2(m^2+z_5)} \, , \nonumber
\\
c_{2,3} &= \frac{1}{2s} \Big[ z_4 (s-m_1^2+m_2^2) + z_5 (s+m_1^2-m_2^2) \nonumber
\\
&\hspace{2em} + s(m_1^2+m_2^2-s) \pm \sqrt{\lambda \, \rho_1} \Big] \, ,
\end{align}
with $\rho_1 = \lambda(s,z_4,z_5) - 4sm^2$. For later convenience we define the polynomial $P_2 \equiv P_2(z_4,z_5,z) \equiv s(z-c_2)(z-c_3)$.

We can now construct the function $\hat{\varphi}^{(1)}(\bm{z})$ with respect to $z$ using Eq.~\eqref{eq:phiz_half_integer}. As an example, we consider linear combinations of the two solutions corresponding to the roots $c_2$ and $c_3$, which gives
\begin{align}
\hat{\varphi}^{(1)}_2(\bm{z}) = \frac{1}{P_2} \frac{\partial P_2 }{\partial z} \frac{\partial P_2}{\partial z_4} - 2\frac{\partial^2 P_2}{\partial z \partial z_4} \, , \quad
\hat{\varphi}^{(1)}_3(\bm{z}) = -\sqrt{\lambda} \sqrt{\rho_1} \frac{1}{P_2} \frac{\partial P_2}{\partial z_4} \, .
\end{align}
For the first solution $\hat{\varphi}^{(1)}_2(\bm{z})$, the remaining variables do not involve square roots, and the construction is straightforward. We finally arrive at
\begin{equation}
\varphi_2(\bm{z}) = \frac{1}{z_1z_2z_3z_4z_5} \left[ \frac{1}{P_2} \frac{\partial P_2 }{\partial z} \frac{\partial P_2}{\partial z_4} - 2\frac{\partial^2 P_2}{\partial z \partial z_4} \right] d^6\bm{z} \, .
\end{equation}
For the second solution $\hat{\varphi}^{(1)}_3(\bm{z})$, noting that $\sqrt{\lambda}$ is independent of $\bm{z}$, we can identify $\Lambda = \rho_1$ for the construction at the level of $z_4$ or $z_5$. The final result is then given by
\begin{equation}
\varphi_3(\bm{z})= -\frac{\sqrt{\lambda}\sqrt{s(s-4m^2)}}{z_1z_2z_3z_4z_5} \frac{1}{P_2} \frac{\partial P_2}{\partial z_4} \, d^6\bm{z} \, .
\end{equation}
The other two solutions can be constructed similarly, which we give in the Supplemental Materials. We have applied the same procedure to all sub-topologies, and hence constructed the full canonical basis for this integral family. The above differential forms can be converted to Feynman integrals using the multivariate intersection theory \cite{Frellesvig:2019uqt, Weinzierl:2020xyy}. As a concrete example, we have
\begin{align}
\bra{\varphi_2} &= \frac{s(s-m_1^2-m_2^2)}{\epsilon} \, \bra{F_{11121}} - \frac{m_2^2(s+m_1^2-m_2^2)}{\epsilon} \, \bra{F_{21111}} \nonumber
\\
&+ \frac{m^2\lambda+sm_1^2m_2^2}{\epsilon^2} \, \bra{F_{21121}} + \frac{2 \big[ 2m^2(s-m_1^2+m_2^2) - sm_2^2 \big]}{\epsilon^2} \, \bra{F_{10221}} \nonumber
\\
&+ \frac{m_2^2}{\epsilon^2} \bra{F_{21002}} + \frac{s}{\epsilon^2} \bra{F_{01220}} - \frac{m_1^2}{\epsilon^2} \bra{F_{21020}} \, ,
\end{align}
where $F_{a_1a_2a_3a_4a_5} \equiv F_{a_1,a_2,a_3,a_4,a_5,0,0}$.
It is then easy to verify that their differential equations are of the $\epsilon$-form. We have also applied our method to the massless and massive double box integrals with success.

\section{Summary and outlook}

To summarize, in this work we have proposed a novel method to construct canonical Feynman integrals using intersection theory. We exploit the fact that Feynman integrals can be expressed as generalized hypergeometric integrals using, e.g., the Baikov parameterization. Such an integral can be regarded as the product of a cocycle and a cycle in the language of twisted (co)homology. The twisted cycle is an equivalence class determined by a multivalued function $u(\bm{z})$ corresponding to a particular integral topology. We then construct all possible cocycles such that the integral takes the $d\log$-form, and project them to Feynman integrals by computing their intersection numbers with a set of arbitrarily chosen master integrals. These then serve as candidates for canonical master integrals from which one can derive differential equations of the $\epsilon$-form.

We have applied our constructive approach to several nontrivial two-loop multi-scale problems.
We find that our method is able to construct all independent $d\log$-forms for the two-loop four-scale triangle integrals. After converting them to Feynman integrals, we have verified that they are indeed canonical ones as expected.
We have also tested our method in the cases of massless and massive double box integrals with success.
Our algorithm can be easily automated and applied to more complicated problems in the future. We emphasize that while we have used the Baikov representation to demonstrate our method, the construction procedure is not confined to that and can be applied to other representations which admit the interpretation as hypergeometric integrals.

It will be of high interests to extend our method to integral families involving elliptic sectors. While a fully canonical basis does not exist in this case, it is very useful to find a basis whose differential equation is ``as canonical as possible''. This has been an active research topic recently (see, e.g., \cite{Weinzierl:2019pfw} and references therein).
Finally, it is well-known that $d\log$-form integrals in planar $\mathcal{N}=4$ supersymmetric theories admit beautiful geometric interpretations \cite{Arkani-Hamed:2013jha, Arkani-Hamed:2016byb}. It is extremely interesting to review the construction of the $d\log$-forms for generic theories from a geometric point of view, which may lead to deeper understanding of the loop amplitudes in these theories.

\vspace{1ex}

This work was supported in part by the National Natural Science Foundation of China under Grant No. 11975030 and 11635001. The research of X. Xu was supported in part by the Swiss National Science Foundation (SNF) under Grant No. $200020\_182038$.

\bibliography{references_inspire,references_local}

\clearpage

\section{Supplemental materials}

\subsection{Canonical bases for one-loop integrals}

Here we give the canonical bases for arbitrary one-loop integrals. A one-loop Feynman integral with $E+1$ external legs has $E+1$ independent propagators which define the integral topology. The Baikov representation for this topology is given by
\begin{align}
F_{a_1,\ldots,a_{E+1}} &= \frac{1}{(4 \pi)^{E/2} \Gamma((d-E)/2)} \int   \frac{\big[ G(\bm{z}) \big]^{(d-E-2)/2}}{\mathcal{K}^{(d-E-1)/2}  }  \prod_{i=1}^{E+1} \frac{dz_i}{z_i^{a_i}} \, ,
\label{eq:1loopbaikov}
\end{align}
where $a_i > 0$ and $\mathcal{K}$ is the Gram determinant of external momenta. It is well-known that at one-loop, there is only one master integral for each topology, therefore we only need to construct one $d\log$-form integral of the type
\begin{equation}
\int_{\mathcal{C}} u(\bm{z}) \varphi(\bm{z}) = \int_{\mathcal{C}}  \bigg[ \frac{G(\bm{z})}{\mathcal{K}}  \bigg]^{-\epsilon} \, F(\bm{z}) \bigwedge_{i=1}^{E+1} \frac{dz_i}{z_i^{a_i}} \, ,
\label{eq:1loopdlog}
\end{equation}
with suitable functions $F(\bm{z})$ and powers $\{a_i\}$.
Comparing Eq.~\eqref{eq:1loopbaikov} to Eqs.~\eqref{eq:baikov_lbl} and \eqref{eq:uz}, we can identify
\begin{equation}
u(\bm{z}) = [G(\bm{z})]^{(2-E)/2-\epsilon} \, \mathcal{K}^{(E-3)/2+\epsilon} \, .
\end{equation}
The $n$-forms $\varphi(\bm{z})$ we'd like to construct takes the form of Eq.~\eqref{eq:phiz}. We may follow the variable-by-variable approach outlined in the main text. However, at one-loop we can actually write down the results directly. If $E$ is even, $\gamma$ is an integer and $[G(\bm{z})]^{\gamma}$ is a rational function. We can then simply choose
\begin{equation}
\varphi(\bm{z}) = \mathcal{K}^{(3-E)/2} \, \big[ G(\bm{z}) \big]^{(E-2)/2} \bigwedge_{i=1}^{E+1} d\log(z_i) \, .
\end{equation}
It is easy to see that the resulting integral is of the $d\log$-form.
On the other hand, if $E$ is odd, $\gamma$ is a half-integer and $[G(\bm{z})]^{\gamma}$ is an algebraic function. In this case we choose
\begin{align}
\varphi(\bm{z}) = \sqrt{G(\bm{0})} \, \mathcal{K}^{(3-E)/2} \, \big[ G(\bm{z}) \big]^{(E-3)/2} \bigwedge_{i=1}^{E+1}  \frac{dz_i}{z_i} \, .
\end{align}
The integrand in Eq.~\eqref{eq:1loopdlog} is then
\begin{align}
u(\bm{z}) \varphi(\bm{z}) &= \mathcal{K}^\epsilon \sqrt{G(\bm{0})} \, \bigwedge_{i=1}^{E+1} \frac{dz_i}{z_i} \big[ G(\bm{z}) \big]^{-1/2-\epsilon} \nonumber
\\
&\hspace{-4em} = \bigg[ \frac{G(\bm{z})}{\mathcal{K}} \bigg]^{-\epsilon} \, \bigwedge_{i=1}^{E+1} \frac{dz_i}{z_i} \sqrt{\frac{G(\vec{0}_i,z_{i+1},\ldots,z_{E+1})}{G(\vec{0}_{i-1},z_{i},\ldots,z_{E+1})}} \, ,
\end{align}
where $\vec{0}_n$ represents $n$ consecutive zeros.
The above product of $E+1$ factors has the property that the $i$-th factor only depends on $z_j$ for $j \geq i$. This property allows us to rewrite it as
\begin{align}
u(\bm{z}) \varphi(\bm{z}) = \bigg[ \frac{G(\bm{z})}{\mathcal{K}} \bigg]^{-\epsilon} \bigwedge_{i=1}^{E+1} d\log f_i(z_i,\ldots,z_{E+1}) \, ,
\end{align}
where the function $f_i$ satisfies
\begin{align}
\frac{\partial}{\partial z_i} \log f_i(z_i,\ldots,z_{E+1}) = \frac{1}{z_i} \sqrt{\frac{G(\vec{0}_i,z_{i+1},\ldots,z_{E+1})}{G(\vec{0}_{i-1},z_{i},\ldots,z_{E+1})}} \, .
\end{align}

We now exploit the fact that the Gram determinants of one-loop integrals are quadratic polynomials of $z_i$. The differential equations satisfied by $f_i$ can then be easily solved using that
\begin{align}
\frac{\partial}{\partial x} \log \frac{1-\sqrt{\frac{x_2(x_1-x)}{x_1(x_2-x)}}}{1+\sqrt{\frac{x_2(x_1-x)}{x_1(x_2-x)}}} = \frac{\sqrt{x_1 x_2}}{x \sqrt{(x_1-x)(x_2-x)}} \, ,
\label{eq:deq_fi_1loop}
\end{align}
up to an irrelevant phase.

After constructing all the $d\log$-form integrals, we now need to convert them to Feynman integrals. This can be achieved using the intersection theory. However, at one loop it turns out to be easier. For the even-$E$ case, the $d\log$-form integral is just the integral $F_{1,\ldots,1}$ in spacetime dimension $E + 2 - 2\epsilon$, i.e.,
\begin{align}
\sqrt{\mathcal{K}} \, F^{(E+2-2\epsilon)}_{1,\ldots,1} = \frac{1}{(4 \pi)^{E/2} \Gamma(1-\epsilon)} \int  \bigg[ \frac{G(\bm{z})}{\mathcal{K}} \bigg]^{-\epsilon} \prod_{i=1}^{E+1}  d\log(z_i) \, .
\end{align}
The above integral can then be expressed by the $d$-dimensional ones via dimensional recurrence relations \cite{Tarasov:1996br}.
Similarly, for the odd-$E$ case, the $d\log$-form integral corresponds to the integral in $E+1-2\epsilon$ dimensions:
\begin{align}
\sqrt{G(\bm{0})} \, F^{(E+1-2\epsilon)}_{1,\ldots,1} &= \frac{1}{(4 \pi)^{E/2} \Gamma(1/2-\epsilon)} \int \mathcal{K}^{\epsilon} \, \sqrt{G(\bm{0})} \, \big[ G(\bm{z}) \big]^{-1/2-\epsilon} \prod_{i=1}^{E+1} \frac{dz_i}{z_i} \, .
\end{align}
Therefore, the canonical basis for one-loop Feynman integrals with arbitrary internal masses and external momenta can be fully constructed using the above procedure. We note that the canonical integrals given here are the same as the $D$-dimensional $D$-gon integrals studied in \cite{Bourjaily:2019exo} (see also \cite{Herrmann:2019upk}).

\subsection{More results of two-loop maximally cut integrals}

Here we give more results for two-loop canonical integrals in the maximally cut case. We first introduce the concept of cuts in the Baikov representation. We consider an integral family defined by $N$ independent propagators. Belonging to this integral family we pick a topology defined by $m$ propagators. Typically we have $m < N$, and to construct the Baikov representation one often needs to introduce more than $m$ Baikov variables $z_i$. Therefore, in general the Baikov representation in the loop-by-loop construction takes the form
\begin{equation}
F_{a_1,\ldots,a_m,0,\ldots,0} = \int_{\mathcal{C}} \bigg[ \prod_{i} \big[ G_i(\bm{z}) \big]^{-\gamma_i-\beta_i\epsilon} \bigg] \bigg[ \prod_{j=1}^{m} \frac{dz_j}{z_j^{a_j}} \bigg] \prod_{k} dz_{k} \, ,
\label{eq:mloop_baikov}
\end{equation}
where the ISPs $z_k$'s are taken from a subset of $\{z_{m+1},\ldots,z_N\}$.

Cutting a Baikov variable $z_j$ ($j \leq m$) for the integral in Eq.~\eqref{eq:mloop_baikov} amounts to changing the integration domain of $z_j$ to an infinitesimal closed contour around the pole $z_j = 0$. The maximally cut version of Eq.~\eqref{eq:mloop_baikov}, where all $z_j$'s ($j=1,\ldots,m$) are cut, is then given by
\begin{equation}
F_{a_1,\ldots,a_m,0,\ldots,0}^{\text{$m$-cut}} = \int_{\mathcal{C}'} \bigg[ \prod_{k} dz_{k} \bigg]
\bigg[ \prod_{j=1}^{m} \oint\limits_{z_j=0} \frac{dz_j}{z_j^{a_j}} \bigg] \times \prod_{i} \big[ G_i(\bm{z}) \big]^{-\gamma_i-\beta_i\epsilon} \, ,
\label{eq:mloop_mcut}
\end{equation}
where the integration domain $\mathcal{C}'$ for the ISPs is determined after integrating out $z_j$ ($j=1,\ldots,m$). The integrations over $z_j$'s can be performed using the residue theorem, giving rise to
\begin{equation}
F_{a_1,\ldots,a_m,0,\ldots,0}^{\text{$m$-cut}} = \int_{\mathcal{C}^\prime} \bigg[ \prod_{k} dz_{k} \bigg] \hat{\varphi}(\bm{z}') \prod_{i} G_{i,0}(\bm{z}')^{-\gamma_i-\beta_i\epsilon} \, ,
\label{eq:mloop_mcut2}
\end{equation}
where $\bm{z}'$ is the collection of the ISPs $\{z_k\}$, $\hat{\varphi}(\bm{z}')$ is a rational function determined by the residues of the integrand in Eq.~\eqref{eq:mloop_mcut} at $z_j = 0$, and
\begin{equation}
G_{i,0}(\bm{z}') \equiv G_i(\bm{z}) \Big|_{z_1=\cdots=z_m=0} \, .
\end{equation}

The maximally cut integrals are of interest on their own. The cut integrals satisfy the same differential equations as the uncut ones \cite{Primo:2016ebd, Primo:2017ipr, Frellesvig:2017aai}. After imposing the maximal cut, all integrals with fewer propagators drop out from the differential equations due to the fact that at least one of the residues at $z_j=0$ vanishes. As a result, only the ``homogeneous'' part of the differential equations remain. Therefore, constructing $d\log$-form integrals in the type of Eq.~\eqref{eq:mloop_mcut2} helps to transform the homogeneous part of the differential equations into the $\epsilon$-form, which serves as the first (and very often the most difficult) step towards a full canonical basis.

After imposing the maximal cuts, there are 3 possibilities: 1) there is no extra ISP left to integrate over; 2) there is exactly one extra ISP left (the univariate case); 3) there are more than one extra ISPs left (the multivariate case). The first case is easy to deal with. There is only one master integral for this top topology, resembling the one-loop case. The homogeneous part of its differential equation can be easily turned into the $\epsilon$-form by multiplying a suitable factor. One can also study the inhomogeneous part by cutting on fewer propagators, leading to the second or the third case. The treatment of the univariate and multivariate cases has been outlined in the main text. In the following, we present two more examples of maximally cut integrals giving rise to the univariate case.

\begin{itemize}

\item{Massless double box.}

The propagators are given by
\begin{multline}
\{ k_1^2, \, (k_1+p_1)^2, \, (k_1+p_1+p_2)^2, \, (k_1+k_2)^2, \, k_2^2, \, (k_2-p_3)^2,
\\
(k_2-p_1-p_2)^2, \, z=(k_2-p_1)^2, \, D_9 \} \, ,
\end{multline}
where $p_i^2=0$, $(p_1+p_2)^2=s$, $(p_1-p_3)^2=t$. We consider the topology $\{1,1,1,1,1,1,1,0,0\}$, for which the 8th propagator $z$ is an ISP in the loop-by-loop Baikov representation. After imposing the maximal cuts, we have
\begin{align}
\mathcal{N}_\epsilon &= \frac{1}{2^6 \pi^3 \Gamma^2(1/2-\epsilon)} \, , \nonumber
\\
u(z) &= \frac{1}{s^2} \bigg( \frac{t(s+t)}{s^2} \bigg)^\epsilon z^{-1-\epsilon} (s+z)^{\epsilon} (t-z)^{-1-2\epsilon} \, , \nonumber
\\
\w &= d\log(u) = \frac{\epsilon}{s+z} dz + \frac{1+2\epsilon}{t-z} dz - \frac{1+\epsilon}{z} dz \, .
\end{align}

The connection $\w$ has two critical points and there is no half-integer coefficient. Therefore according to  Eq.~\eqref{eq:phiz_integer}, we can construct two 1-forms
\begin{equation}
\phi_1 = s^2 z dz \, , \quad  \phi_2 = s^2 (t-z) dz \, .
\end{equation}
We denote their corresponding Feynman integrals as $I_1$ and $I_2$. To find their expressions, we pick the basis as $E_1=F_{1,1,1,1,1,1,1,0,0}$ and $E_2=F_{1,2,1,1,1,1,1,0,0}$. Their corresponding cocycles are
\begin{equation}
\bra{e_1} = dz \, , \quad \bra{e_2}= \frac{1+2\epsilon}{z} dz \, .
\end{equation}
We can then perform the decomposition which gives
\begin{align}
I_1 &= -\frac{s(1+3\epsilon)}{2\epsilon} E_1 + \frac{st(1+\epsilon)}{2\epsilon(1+2\epsilon)} E_2 \, , \nonumber
\\
I_2 &= \frac{s(1+3\epsilon) + 2\epsilon t}{2\epsilon} E_1 - \frac{s t (1+\epsilon)}{2\epsilon(1+2\epsilon)} E_2 \, .
\end{align}
We emphasize that the above expressions are valid at the level of maximal cuts, i.e., on the right hand side there are more contributions from sub-topologies.

The differential equations of $I_1$ and $I_2$ (trimming sub-topologies) with respect to $s$ and $t$ are given by
\begin{align}
\frac{\partial}{\partial s}
\begin{pmatrix}
I_1
\\
I_2
\end{pmatrix}
&= \epsilon
\begin{pmatrix}
-\frac{2}{s} & \frac{1}{s+t}
\\
\frac{2}{s} & -\frac{s+2t}{s(s+t)}
\end{pmatrix}
\begin{pmatrix}
I_1
\\
I_2
\end{pmatrix}
\, , \nonumber
\\
\frac{\partial}{\partial t}
\begin{pmatrix}
I_1
\\
I_2
\end{pmatrix}
&= \epsilon
\begin{pmatrix}
0 & -\frac{s}{t (s+t)}
\\
-\frac{2}{t} & -\frac{s}{t (s+t)}
\end{pmatrix}
\begin{pmatrix}
I_1
\\
I_2
\end{pmatrix}
\, .
\end{align}
One can see that the differential equations are indeed of the $\epsilon$-form. 

\item{Internally massive double box.}

The propagators are
\begin{multline}
\{ k_1^2, \, (k_1+p_1)^2, \, (k_1+p_1+p_2)^2, \, (k_1+k_2)^2-m^2, \, k_2^2-m^2, \, (k_2-p_3)^2-m^2,
\\
(k_2-p_1-p_2)^2-m^2, \, z=(k_2-p_1)^2-m^2, \, D_9 \} \, ,
\end{multline}
with $p_i^2=0$, $(p_1+p_2)^2=s$, $(p_1-p_3)^2=t$. We again consider the maximally cut integrals in the topology $\{1,1,1,1,1,1,1,0,0\}$, and we have
\begin{align}
\mathcal{N}_\epsilon &= \frac{1}{2^6 \pi^3  \Gamma^2(1/2-\epsilon)} \, , \nonumber
\\
u(z) &= \frac{1}{s^2} \bigg( \frac{t(s+t)}{s^2} \bigg)^\epsilon z^{-1-2\epsilon} \big[ (z-c_0) (z-c_1) \big]^{-1/2-\epsilon} \big[ (z-c_2) (z-c_3) \big]^{\epsilon}  \, , \nonumber
\\
\w &= d\log(u) = - \frac{1+2\epsilon}{z} - \frac{1/2+\epsilon}{z-c_0} - \frac{1/2+ \epsilon}{z-c_1} + \frac{\epsilon}{z-c_2} + \frac{\epsilon}{z-c_3} \, ,
\end{align}
with
\begin{align}
c_{0,1} &= \frac{s t \pm 2 \sqrt{m^2 s t (s+t)}}{s} \, , \nonumber
\\
c_{2,3} &= \frac{1}{2} \Big[ -s \pm \sqrt{s (s-4m^2)} \Big] \, .
\end{align}

According to Eq.~\eqref{eq:phiz_half_integer}, we construct the 1-forms as
\begin{align}
\phi_1(z) &= s^2 \sqrt{c_0 c_1} dz \, , \quad \phi_4(z) = s^2z dz \, , \nonumber
\\
\phi_{2,3}(z) &= \frac{s^2 z \sqrt{(c_0-c_{2,3})(c_1-c_{2,3})}}{z-c_{2,3}} dz \, .
\end{align}
Again we have verified that the homogeneous part of the corresponding differential equations is of the $\epsilon$-form.

\end{itemize}

\subsection{The complete canonical basis for the two-loop four-scale triangle integrals}

Here we give the complete results for the two-loop four-scale triangle integrals in the sector $\{1,1,1,1,1,0,0\}$. We construct the loop-by-loop Baikov representation in the order $\{k_1,k_2\}$. The polynomials entering the $u$ function are given by
\begin{align}
P_1(z_5,z) &\equiv -4G(k_2,p_2) \equiv (z-c_0)(z-c_1) \, , \nonumber
\\
P_2(z_4,z_5,z) &\equiv 4G(k_2,p_1,p_2) \equiv s(z-c_2)(z-c_3) \, , \nonumber
\\
P_3(z_1,z_2,z_3,z_5,z) &\equiv 4G(k_1,k_2,p_2) \equiv (z_3+m^2)(z-c_4)(z-c_5) \, ,
\end{align}
where the roots are
\begin{align}
c_{0,1} &= m_2^2 + z_5 \pm 2 \sqrt{m_2^2(m^2+z_5)} \, , \nonumber
\\
c_{2,3} &= \frac{1}{2s} \Big[ z_4 (s-m_1^2+m_2^2) + z_5 (s+m_1^2-m_2^2) + s (m_1^2+m_2^2-s) \pm \sqrt{\lambda \, \rho_1} \Big] \, , \nonumber
\\
c_{4,5} &= \frac{1}{2(z_3+m^2)} \Big[ m_2^2 (z_2+z_3-z_5) + z_5 (2m^2+z_1+z_3) \nonumber
\\
&\hspace{8em} + (z_1-z_3) (2m^2-z_2+z_3) \pm \sqrt{\rho_2 \rho_3} \Big] \, ,
\end{align}
with
\begin{align}
\rho_1 &\equiv \lambda(s,z_4,z_5) - 4sm^2 \, , \quad
\rho_2 \equiv \lambda(m_2^2,z_1,z_3) - 4 m_2^2 m^2 \, , \quad
\rho_3 \equiv \lambda(z_2,z_3,z_5) - 4 z_2 m^2 \, , \nonumber
\\
\lambda &\equiv \lambda(s,m_1^2,m_2^2) \, , \quad \lambda(x,y,z) \equiv x^2 + y^2 + z^2 - 2xy - 2yz - 2zx \, .
\end{align}
We then have for the Baikov representation
\begin{align}
\mathcal{N}_\epsilon &= \frac{(-\lambda)^\epsilon}{4 \pi^2 \Gamma^2(1-\epsilon)} \, , \nonumber
\\
u(\bm{z}) &= \frac{(-\lambda)^{-1/2}}{\left[ -P_1(z_5,z) \right]^{1/2-\epsilon}} \big[ P_2(z_4,z_5,z) \big]^{-\epsilon} \big[ P_3(z_1,z_2,z_3,z_5,z) \big]^{-\epsilon} \nonumber
\\
&= \frac{(-\lambda)^{-1/2} \, s^{-\epsilon} (z_3+m^2)^{-\epsilon}}{\left[ -(z-c_0)(z-c_1) \right]^{1/2-\epsilon}} \prod_{i=2}^{5} (z-c_i)^{-\epsilon} \, .
\end{align}

We now pick $z$ as the first variable and perform the construction according to Eq.~\eqref{eq:phiz_half_integer}. The first solution is simply
\begin{equation}
\hat{\phi}^{(1)}_1(\bm{z}) = \sqrt{\lambda} \, .
\end{equation}
The construction for the remaining variables is trivial and we arrive at
\begin{align}
\varphi_1(\bm{z}) = \frac{\sqrt{\lambda} \, d^6\bm{z}}{z_1 z_2 z_3 z_4 z_5} \, .
\end{align}

The second and third solutions in Eq.~\eqref{eq:phiz_half_integer} correspond to the roots $c_2$ and $c_3$:
\begin{align}
\hat{\phi}^{(1)}_2(\bm{z}) &=\sqrt{\lambda} \frac{\sqrt{(c_0-c_2)(c_1-c_2)}}{z-c_2} =\sqrt{\lambda} \frac{\sqrt{\lambda} (s-z_4+z_5) - \sqrt{\rho_1} (s-m_1^2+m_2^2)}{2 s (z-c_2)} \, , \nonumber
\\
\hat{\phi}^{(1)}_3(\bm{z}) &=\sqrt{\lambda}\frac{\sqrt{(c_0-c_3)(c_1-c_3)}}{z-c_3} = \sqrt{\lambda}\frac{\sqrt{\lambda} (s-z_4+z_5) + \sqrt{\rho_1} (s-m_1^2+m_2^2)}{2 s (z-c_3)} \, .
\end{align}
It will be instructive to write them in a different way
\begin{align}
\hat{\phi}^{(1)}_2(\bm{z}) &= \left[ \frac{1}{(z-c_{2})} \frac{\partial}{\partial z_4} - \frac{\partial^2}{\partial z_4 \partial z} \right] P_2(z_4,z_5,z) \, , \nonumber
\\
\hat{\phi}^{(1)}_3(\bm{z}) &= \bigg[ \frac{1}{(z-c_{3})} \frac{\partial}{\partial z_4} - \frac{\partial^2}{\partial z_4 \partial z} \bigg] P_2(z_4,z_5,z) \, .
\end{align}
The above expressions are not rational functions of $z_i$ due to the appearance of $\sqrt{\rho_1}$ in $c_2$ and $c_3$. To perform the construction for the remaining variables, we need to take linear combinations of them such that $\sqrt{\rho_1}$ either disappears, or becomes an overall factor. Observing that $c_i$'s are roots of quadratic polynomials, we know for example that $(z-c_2)(z-c_3)$ is a rational function. Therefore we can take the following combinations
\begin{align}
\hat{\varphi}^{(1)}_2(\bm{z}) &= \hat{\phi}^{(1)}_2(\bm{z}) + \hat{\phi}^{(1)}_3(\bm{z}) = \frac{1}{P_2} \frac{\partial P_2 }{\partial z} \frac{\partial P_2}{\partial z_4} - 2\frac{\partial^2 P_2}{\partial z \partial z_4} \, ,\nonumber
\\
\hat{\varphi}^{(1)}_3(\bm{z}) &= \hat{\phi}^{(1)}_3(\bm{z}) - \hat{\phi}^{(1)}_2(\bm{z}) = -\sqrt{\lambda} \sqrt{\rho_1} \frac{1}{P_2} \frac{\partial P_2}{\partial z_4} \, .
\end{align}

We now note that $\hat{\varphi}^{(1)}_i(\bm{z})$ takes the general form $\sqrt{\Lambda(\bm{z}')} g(\bm{z})$, where $\Lambda(\bm{z}')$ and $g(\bm{z})$ are rational functions of the variables, with $\bm{z}' = \{z_1,\ldots,z_5\}$. To make the final 6-forms rational in all variables, we need to take care of these $\sqrt{\Lambda(\bm{z}')}$ factors in the remaining construction. For $\hat{\varphi}^{(1)}_2(\bm{z})$, there are no algebraic factors we need to consider, and the complete result is simply
\begin{align}
\varphi_2(\bm{z}) &= \frac{1}{z_1z_2z_3z_4z_5} \left[ \frac{1}{P_2} \frac{\partial P_2 }{\partial z} \frac{\partial P_2}{\partial z_4} - 2\frac{\partial^2 P_2}{\partial z \partial z_4} \right] d^6\bm{z} \, .
\end{align}
For $\hat{\varphi}^{(1)}_3(\bm{z})$, since $\sqrt{\lambda}$ is independent of the $\bm{z}$ variables, we can simply treat it as a constant factor and take $\Lambda = \rho_1$. Since $\rho_1$ only depends on $z_4$ and $z_5$, the construction for $z_1$, $z_2$ and $z_3$ is straightforward. For $z_4$ and $z_5$, we may use Eq.~\eqref{eq:deq_fi_1loop}. The result is then
\begin{align}
\varphi_3(\bm{z}) &= -\frac{\sqrt{\lambda}\sqrt{s(s-4m^2)}}{z_1z_2z_3z_4z_5} \frac{1}{P_2} \frac{\partial P_2}{\partial z_4} \, d^6\bm{z} \, .
\end{align}

Using the same method with the two roots $c_4$ and $c_5$, we can obtain the other two solutions
\begin{align}
\hat{\phi}^{(1)}_4(\bm{z}) &= \frac{\sqrt{\lambda}}{\sqrt{\rho_2}} \bigg[ \frac{1}{(z-c_{4})} \frac{\partial}{\partial z_2} - \frac{\partial^2}{\partial z_2 \partial z} \bigg] P_3(z_1,z_2,z_3,z_5,z) \, ,\nonumber
\\
\hat{\phi}^{(1)}_5(\bm{z}) &= \frac{\sqrt{\lambda}}{\sqrt{\rho_2}}  \bigg[ \frac{1}{(z-c_{5})} \frac{\partial}{\partial z_2} - \frac{\partial^2}{\partial z_2 \partial z} \bigg] P_3(z_1,z_2,z_3,z_5,z) \, .
\end{align}
Taking the linear combination with a plus gives
\begin{align}
\hat{\varphi}^{(1)}_4(\bm{z}) &= \hat{\phi}^{(1)}_4(\bm{z}) + \hat{\phi}^{(1)}_5(\bm{z}) = \frac{\sqrt{\lambda}}{\sqrt{\rho_2}} \left[ \frac{1}{P_3} \frac{\partial P_3 }{\partial z} \frac{\partial P_3}{\partial z_2} - 2\frac{\partial^2 P_3}{\partial z \partial z_2} \right] . \nonumber
\end{align}
Starting from $\hat{\varphi}^{(1)}_4(\bm{z})$ we can continue the construction which gives the final result
\begin{align}
\varphi_4(\bm{z}) &=  \frac{\sqrt{\lambda}\sqrt{m_2^2(m_2^2-4m^2)}}{z_1z_2z_3z_4z_5} \frac{1}{\rho_2} \left[ \frac{1}{P_3} \frac{\partial P_3 }{\partial z} \frac{\partial P_3}{\partial z_2} - 2\frac{\partial^2 P_3}{\partial z \partial z_2} \right] d^6\bm{z} \, .
\end{align}
On the other hand, we find that the other linear combination with a minus sign does not lead to an independent integral for this top topology. This is expected since there are only 4 master integrals for the top topology, in accordance with the results in the maximally cut case. We note that this additional linear combination actually belongs to a sub-topology.

The above 4 $d\log$-form integrals can be projected to Feynman integrals using intersection theory, which gives
\begin{align}
\bra{\varphi_1} &= \sqrt{\lambda} \, \bra{F_{11111}} \, , \nonumber
\\
\bra{\varphi_2} &= \frac{s(s-m_1^2-m_2^2)}{\epsilon} \, \bra{F_{11121}} - \frac{m_2^2(s+m_1^2-m_2^2)}{\epsilon} \, \bra{F_{21111}} \nonumber
\\
&+ \frac{m^2\lambda+sm_1^2m_2^2}{\epsilon^2} \, \bra{F_{21121}} + \frac{2 \big[ 2m^2(s-m_1^2+m_2^2) - sm_2^2 \big]}{\epsilon^2} \, \bra{F_{10221}} \nonumber
\\
&+ \frac{m_2^2}{\epsilon^2} \bra{F_{21002}} + \frac{s}{\epsilon^2} \bra{F_{01220}} - \frac{m_1^2}{\epsilon^2} \bra{F_{21020}}  \, , \nonumber
\\
\bra{\varphi_3} &= \frac{\sqrt{\lambda} \sqrt{s(s-4m^2)}}{\epsilon} \, \bra{F_{11121}} \, , \nonumber
\\
\bra{\varphi_4} &= \frac{\sqrt{\lambda} \sqrt{m_2^2(m_2^2-4m^2)}}{\epsilon} \, \bra{F_{21111}} \, .
\end{align}
where $F_{a_1a_2a_3a_4a_5} \equiv F_{a_1,a_2,a_3,a_4,a_5,0,0}$.

Using the same procedure, one may also construct $d\log$-form integrals which correspond to loop integrals in sub-topologies. This can be done by building functions $\hat{\varphi}(\bm{z})$ with reduced number of propagators in the denominator.
Alternatively, we can start from the $u(\bm{z})$ function for each sub-topology, and perform the construction accordingly. In the end, we are able to construct all canonical master integrals for all integral families in the two-loop four-scale triangle diagrams.

\end{document}